\documentclass{elsarticle}

\usepackage{epic}

\usepackage{latexsym}
\usepackage{amsmath}
\usepackage{epsfig}
\usepackage{amssymb}
\usepackage{enumerate}
\usepackage{natbib}
\usepackage[mathscr]{euscript}

\newtheorem{theorem}{Theorem}

\newtheorem{proposition}{Proposition}

\newtheorem{claim}{Claim}

\newcommand{\loglike}{\mathscr{L}}
\newcommand{\E}{\mathbb{E}}
\renewcommand{\P}{\mathbb{P}}
\newcommand{\ind}{\mathbf{1}}
\newcommand{\allchi}{\mathcal{X}}
\newcommand{\pars}{\mathscr{P}}
\newcommand{\species}{\mathcal{T}}
\newcommand{\deepest}{\mathcal{D}}
\newcommand{\eps}{\varepsilon}
\newcommand{\esf}{\mathrm{ESF}}
\renewcommand{\Xi}{\mathcal{C}}

\begin{document}
\begin{frontmatter}

\title{Likelihood-based tree reconstruction on a concatenation of alignments can be positively misleading}
\author{Sebastien Roch$^{a}$ and Mike Steel$^{b}$}
\address{$^{a}$ Department of Mathematics, University of Wisconsin--Madison, Madison WI, USA\\
$^{b}$ MS Biomathematics Research Centre, University of Canterbury, Christchurch, New Zealand}

\date{\today}

\begin{abstract}
The reconstruction of a species tree from genomic data faces a double hurdle. First, the (gene) tree describing the evolution of each gene may differ from the species tree, for instance, due to incomplete lineage sorting.
Second, the aligned genetic sequences at the leaves of each gene tree provide merely an imperfect estimate of the topology of the gene tree. In this note, we demonstrate formally that a basic statistical problem arises if one tries to avoid accounting for these two processes and analyses the genetic data directly via a concatenation approach.  More precisely, we show that, under the multi-species coalescent with a standard site substitution model, 
maximum likelihood estimation on sequence data that has been concatenated across genes and performed under the incorrect
assumption that all sites have evolved independently and identically
on a fixed tree
is a statistically inconsistent estimator of the species tree. 
Our results provide a formal justification of simulation results described of Kubatko and Degnan (2007) and others, 
and complements recent theoretical results 
by DeGorgio and Degnan (2010) and Chifman and Kubtako (2014).  
\end{abstract}

\begin{keyword}
Phylogenetic reconstruction, 
incomplete lineage sorting,
maximum likelihood,
consistency
\end{keyword}

\end{frontmatter}





\newpage

\section{Introduction}
Modern molecular sequencing technology has provided a wealth of data to help biologists infer evolutionary relationships between species.  Not only is it possible to quickly sequence a single gene across a wide range of species, but hundreds, or even thousands of genes can also be sequenced across those taxa. But with this abundance of data comes new statistical and mathematical challenges. These arise because tree inference requires dealing with the interplay of two random processes, as we now explain.

For each gene, the associated aligned sequence data provides an estimate of the evolutionary {\em gene tree} that  describes the ancestry of this gene as one traces back its ancestry  in time (each copy being inherited from one parent in the previous generation).   Moreover, given sufficiently long sequences, several methods (e.g. maximum likelihood and corrected distance methods) have been shown to be statically consistent estimators of the gene tree topology under various site substitution models   \citep{fel04}.  `Statistical consistency' here refers to the usual  notion in molecular phylogenetics, namely that  as the sequence length grows, the probability that the correct gene tree topology is returned from the data converges to 1 as the number of sites grow.  Here the site patterns generated independently and identically (i.i.d.) under the substitution model on a binary (fully-resolved) gene tree.

But inferring a gene tree is only part of the puzzle of reconstructing the main evolutionary object of interest in biology -- namely a {\em species tree}.   This latter tree describes, on a broad (macroevolutionary) scale, how lineages (consisting of populations of a species) successively separated  and diverged from each other over evolutionary time scales,  with some lineages forming  new species, ultimately leading to the given taxa observed at the present  (a precise definition of a species-level phylogenetic tree is problematic as it requires first agreeing on a definition of `species', for which there are multitude of differing opinions) \citep{mad97,  nic01}.   A species tree, together with the length (time-scale) and width (population size) of its branches, induces a probability distribution on the possible gene trees and, when the discordance between gene trees
is attributed to incomplete lineage sorting, this probability distribution can be described by the so-called  {\em multi-species coalescent} process (details are provided in the recent book  by \cite{kno10}).  This process extends the celebrated {\em Kingman coalescent} process from a single population to a phylogenetic tree, where the latter can be viewed as a `tree of populations' 

The relationship between gene trees and species trees has attracted a good deal of attention from mathematicians and statisticians  over the last decade or so \citep{deg09a, hua10, liu09a, liu09b, roc13, ros02}.   An early and easily verified result  is that for three taxa, the most probable gene tree topology under the multi-species coalescent matches the species tree (the other two competing binary topologies have equal but lower probability) \cite{taj83}.  Consequently, estimating the species tree by the gene tree that appears most frequently is a statistically consistent method (under the multi-species coalescent) when we have just three taxa. Moreover, when there are more than three taxa, one can still estimate a species tree consistently, for example, by estimating all the rooted triples, and using these to reconstruct the species tree topology \citep{deg09b}.  

However, the alternative simple `majority rule' strategy of estimating 
the species tree by merely taking the most frequent gene tree falls apart when we have more than than three species. With four taxa, the most probable gene tree topology can differ from  certain (unbalanced) species tree topologies, while for five or more taxa a more striking result applies -- {\em every} species tree topology has branch lengths for which the most probable gene tree topology differs from that of the species tree (for details, see \cite{deg09a}).    Nevertheless one can still infer a species tree in a statistically consistent manner from a series of gene trees generated i.i.d.~by the multi-species coalescent process, and several techniques have been developed for this (see e.g. \cite{mos10}).  There are also additional mechanisms that can lead to conflict between gene trees and species trees, including reticulate evolution (e.g. the formation of hybrid species), lateral gene transfer (in prokaryotic taxa such as bacteria) and gene duplication and loss, but we do not consider these processes here.  

We have so far discussed these two random processes -- the evolution of sequence site patterns on a gene tree under a site-substitution model, and the random generation of gene trees from the species tree under the multi-species coalescent process -- as separate process. But in reality these two processes work in concert, a gene tree will have a random topology (determined by the multi-species coalescent on the species tree) and on this random gene tree sequences will evolve according to a substitution process.  Thus, it is not immediately obvious whether methods exist for inferring a species tree topology directly from a series of aligned sequences (one for each gene) which would be  statistically consistent as the number of genes grows.  Using techniques from algebraic statistics,  \citet{chi14} recently established that the species tree topology (up to the placement of the root) is an identifiable discrete parameter under the combined substitution--coalescence process.  Moreover they describe an explicit method for estimating the species tree based on phylogenetic invariants and singular value decomposition techniques.
For  Bayesian inference of species trees directly  from sequence data (e.g. via the program *BEAST, \citep{hel10}) the statistical consistency has also been formally established \citep{ste13}. 

In this paper we consider a simpler and alternative strategy that has been used widely for inferring the species tree directly from sequence data, namely concatenation of sequences (e.g.~\citep{mer11,rok03}).
In its simplest form, this strategy simply concatenates all the sequences, and treats them as though each site had evolved i.i.d.~on a fixed tree.
\citet{kub07} used simulations to  study the performance of such a concatenation approach, and their finding suggested that it could lead to misleading phylogenetic estimates. 
Nevertheless, the accuracy of concatenation methods is still 
very much under debate
(e.g.~\citep{gat13,son12,wus13}).
While many simulation studies have concluded that concatenation
methods are significantly less accurate than ILS-based methods
or are prone to producing erroneous estimates with high confidence~\citep{hel10,kub07,kub09,lar10,lea11},
others have found that they can be more accurate
under some conditions (such as low phylogenetic 
signal)~\citep{bay13,gad05,mir14}.  
Moreover, a formal proof of whether or not a standard statistical method, such as maximum likelihood (ML), is statistically consistent as an estimator of tree topology based on
concatenated sequences has never been presented,
with the exception of the work of~\citet{deg10}
who established the consistency of ML in the 
special case of three taxa under
a molecular clock. 

This is the motivation for our current paper.  We consider what happens when ML  is applied under the assumption that the sites evolve i.i.d.~on a fixed tree (in keeping with the concatenation approach).   Our main result (Theorem 1) shows that ML is statistically inconsistent as an estimator of tree topology.   Indeed the probability that the true species tree is an ML tree can be made as small as we wish in the limit as the number of genes grows (even with six taxa).  What makes this result non-trivial is that  studying the behavior of mis-specified 
likelihoods can be challenging.  Our proof of inconsistency involves combining a number of arguments and results, including a classic result in populations genetics (the `Ewens' Sampling formula'), a formal linkage between likelihood and parsimony, and the interplay of various concentration and approximations bounds.

\section{Definitions and main result}

Consider:

\begin{itemize} 
\item a species tree topology $T$ together with  branch lengths $L$  (which, for  each edge $e$ of $T$, combine temporal branch lengths ($t_e$) and an effective population size
for that edge $N_e$ -- note the subscript $e$ here refers to the edge $e$ not `effective'). 

\item $g$  alignments $A_1, A_2, \ldots, A_g$, where $A_i$ consists
of sequences of length $\ell = \ell(g)$ evolved i.i.d. under a symmetric $r$-state site substitution model at substitution rate $\theta$ on the random gene  tree (with associated branch lengths) that is  generated by $(T, L)$ via the multispecies coalescent model.  That is, on each branch of $T$, 
looking backwards in time, lineages entering the branch coalesce
at constant rate according to the Kingman coalescent with fixed
population size. The remaining lineages at the top of the branch
enter the ancestral population. For each locus, conditioned on the generated gene tree,
the alignment is generated according to the symmetric
$r$-state model.

\item 
maximum likelihood tree(s) $T_{ML}$ for the concatenated alignment $A_1A_2 \cdots A_g$ inferred under the assumption that all sites evolve i.i.d. on a tree according to the symmetric $r$-state site substitution model (for branch lengths that are optimized, as usual, as part of  the ML estimation).
 \end{itemize}
 
 Let  $P(T, L, r, g, \ell, \theta)$ be the probability that  $T$ has the same unrooted topology as (at least one) ML tree $T_{ML}$.  Our main result can be stated as follows.

\begin{theorem}
\label{main}
Under the model described above, there exist tree topologies $T$ with branch lengths $L$ for $T$, and a site substitution rate $\theta$ sufficiently small, for which the following holds:  For any $\delta>0$, there is a value $g_0$ so that  
$$P(T, L, r, g, \ell, \theta) \leq \delta$$
for all $g \geq g_0$, and for all 
sequence length functions $\ell = \ell(g)$.
\end{theorem}

\section{Heuristic argument and a key preliminary result}
\label{section:preliminaries}

The formal proof of  Theorem~\ref{main} is presented in the next section. Here we describe the idea of the proof, and establish a
preliminary result that is central to the proof.
 
 In the anomaly zone, the most
frequent gene tree topology differs from the species tree topology.
That in itself does not imply that maximum likelihood
on the concatenation will pick the wrong tree. 
However what we show is that the wrong topology does indeed
lead to a higher expected likelihood. 
We exploit a connection
to parsimony: at low mutation rates
the likelihood score is roughly equal to the
parsimony score (up to a factor). 
The latter, being combinatorial in nature,
turns out to be easier to characterize.
In particular, we show that under the multispecies
coalescent the wrong topology 
has a higher expected parsimony score. 
The following 
preliminary result (Proposition \ref{TT'}) establishes the 
previous claim under the related infinite-alleles model
of mutation.  This proposition plays a key role in the final step (Claim \ref{claim:final}) of the proof of the theorem. 

Given allele frequencies $(a_1, a_2, \ldots, a_n)$ where $\sum_{j=1}^n ja_j =n$, the celebrated `Ewens' Samping Formula' describes the probability of generating such an allele distribution in a coalescent tree, with scaled mutation rate $\theta = 4N\mu$ under an infinite alleles model:

$$P_{\theta, n}(a_1, a_2, \ldots, a_n) = \frac{n!}{\theta_{(n)}}\prod_{j=1}^n\frac{ \left(\theta/j\right)^{a_j}}{a_j!},$$
where $\theta_{(n)} = \theta(\theta +1) \cdots (\theta + n-1).$
(for details, see \citet{dur08}, p.18).  We will apply this in the current setting, where $n=6$ and $\theta = \epsilon$ a small positive constant (to be determined later).

Let ${\bf x} = (0,1,0,1,0,0)$ and ${\bf y} = (0,0,2,0,0,0).$  Then

\begin{equation}
\label{Px}
P_{\epsilon, 6}({\bf x}) = \frac{6!}{\epsilon_{(6)}} \frac{(\epsilon/2)^1}{1!}\frac{(\epsilon/4)^1}{1!} = \frac{3}{4}\epsilon + O(\epsilon^2).
\end{equation}
Similarly,
\begin{equation}
\label{Py}
P_{\epsilon, 6}({\bf y}) = \frac{6!}{\epsilon_{(6)}} \frac{(\epsilon/3)^2}{2!}= \frac{1}{3}\epsilon + O(\epsilon^2).
\end{equation}

\begin{figure}[htb]
\centering
\includegraphics[scale=0.3]{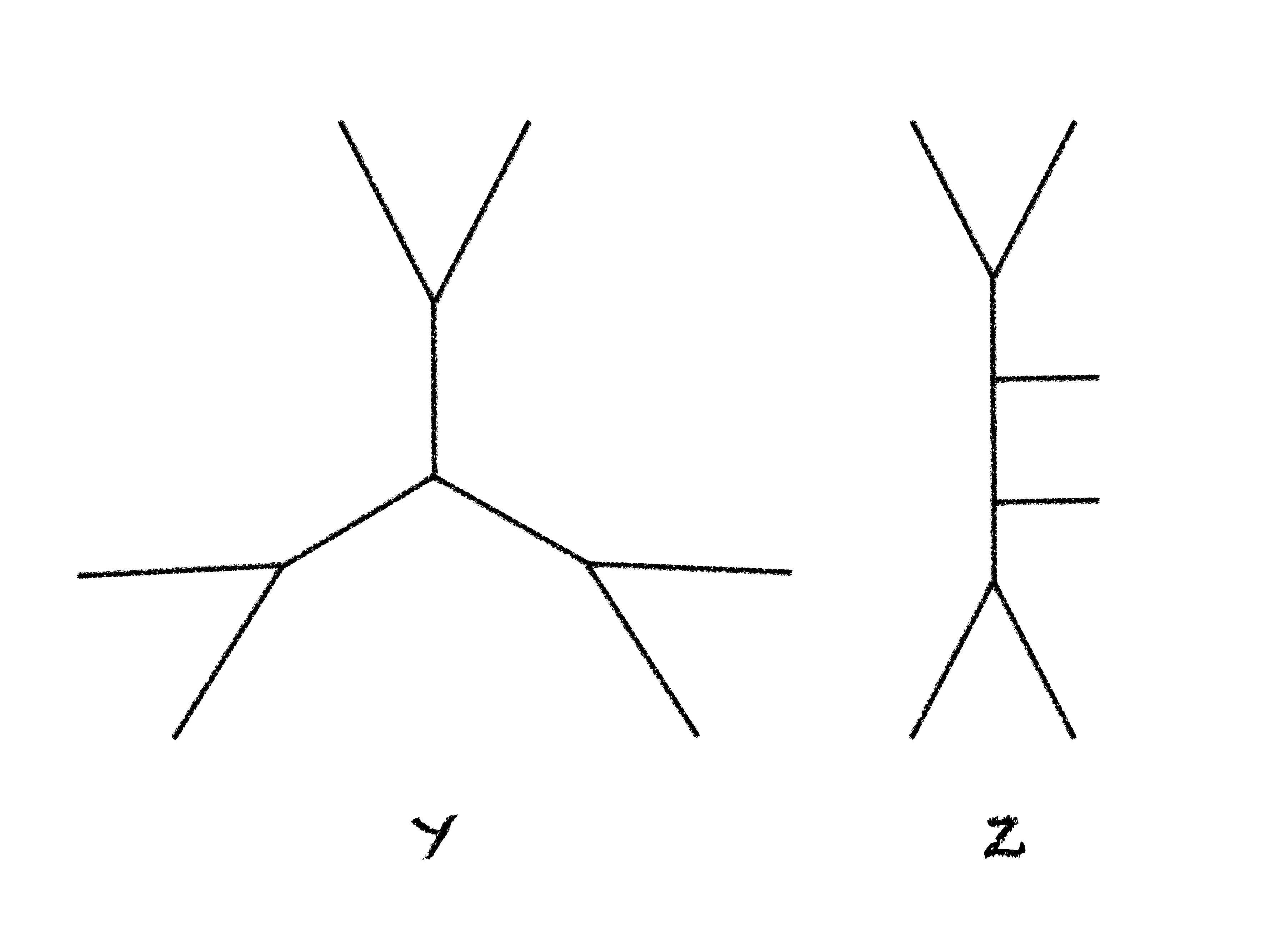}
\caption{The two binary tree shapes on six leaves: (a) the shape $Y$; (b) the shape $Z$.  There are 15 and 90 phylogenetic trees on a given leaf set that have the shapes $Y$ and $Z$, respectively.}
\label{fig1}
\end{figure}

Consider the two unrooted binary tree shapes on six leaves, shown in Fig.~\ref{fig1}, and denote these as $Y$ (the symmetric tree with three cherries) and $Z$ (the caterpillar tree with two cherries).

We apply the above calculations to establish the following result.

\begin{proposition}
\label{TT'}
Let $T$ and $T'$ be two unrooted binary phylogenetic trees of shapes $Z$ and $Y$ respectively.
Consider a site pattern that is  randomly generated on a coalescent tree on the same leaf set under the infinite alleles model
with scaled mutation rate $\theta (= 4N\mu)= \epsilon$.  
For a binary tree topology $W$,
$\pars_W(\chi)$ denote the parsimony score of a site pattern 
$\chi$ on $W$. Then
$$\E_{\esf}[\pars_Z(\chi) - \pars_Y(\chi)] = \frac{1}{60}\epsilon + O(\epsilon^2).$$
where $\E_{\esf}$ denotes the expectation under the infinite-alleles
model. 
\end{proposition}
{\em Proof:}
We refer to a binary pattern on the leaf set $\{1,2,3,4,5,6\}$  as a {\em $k$-clade} if there are $k$ leaves in one state, and $6-k$ in another ($k \leq n/2$).
Given such a binary pattern, the {\em additional penalty} of this clade is its homoplasy score (i.e. the parsimony score minus 1, unless
the clade is a $0$-clade in which case the penalty is 0).

For a phylogenetic tree having shape $Y$ there are:
\begin{itemize}
\item  $\binom{3}{2} \cdot 2 \cdot 2 = 12$ in total 2-clades that cost an additional penalty of $+1$;
\item $\frac{1}{2}\binom{3}{1}\cdot \binom{2}{1} \cdot 2 =6$ in total  3-clades that cost an additional penalty of $+1$;
\item $\frac{1}{2}2 \cdot 2 \cdot 2 = 4$ in total 3-clades that cost an additional penalty of $+2$. 
\end{itemize}

Thus the expected value of the additional parsimony penalty $\Delta_Y$ for a tree phylogenetic tree having shape $Y$ is:

$$1 \cdot P_{\epsilon, 6}({\bf x}) \cdot \frac{12}{\binom{6}{2}} + 1\cdot P_{\epsilon, 6}({\bf y})\cdot \frac{6}{\frac{1}{2}\binom{6}{3}}+ 2 \cdot P_{\epsilon, 6}({\bf y})\cdot \frac{4}{\frac{1}{2}\binom{6}{3}}.$$
Substituting Eqns.~(\ref{Px}) and (\ref{Py}) into this last expression gives:
\begin{equation}
\label{ee1}
\E_{\esf}[\pars_Y(\chi)] = \frac{16}{15} \epsilon + O(\epsilon^2).
\end{equation}

A similar analysis for a $Z$-shape tree shows that there are:

\begin{itemize}
\item $13$ in total 2-clades that cost an additional penalty of $+1$;
\item$5$ in total  3-clades that cost an additional penalty of $+1$;
\item $4$ in total 3-clades that cost an additional penalty of $+2$. 
\end{itemize}

Thus the expected value of the additional parsimony penalty $\pars_Z(\chi)$ for a tree phylogenetic tree having shape $Z$ is:

$$1 \cdot P_{\epsilon, 6}({\bf x}) \cdot \frac{13}{\binom{6}{2}} + 1\cdot P_{\epsilon, 6}({\bf y})\cdot \frac{5}{\frac{1}{2}\binom{6}{3}}+ 2 \cdot P_{\epsilon, 6}({\bf y})\cdot \frac{4}{\frac{1}{2}\binom{6}{3}}.$$
Substituting Eqns.~(\ref{Px}) and (\ref{Py}) into this last expression gives:
\begin{equation}
\label{ee2}
\E_{\esf}[\pars_Z(\chi)] = \frac{13}{12} \epsilon + O(\epsilon^2).
\end{equation}

Combining Eqns.~(\ref{ee1}) and ~(\ref{ee2}) gives:
\begin{equation}
\label{eps}
\E_{\esf}[\pars_Z(\chi) - \pars_Y(\chi)] = \frac{1}{60}\epsilon + O(\epsilon^2).
\end{equation}

Proposition~\ref{TT'} now follows from (\ref{eps}). 

\hfill$\Box$

\section{Proof of Theorem~\ref{main}}

\begin{figure}[htb]
\centering
\includegraphics[scale=0.3]{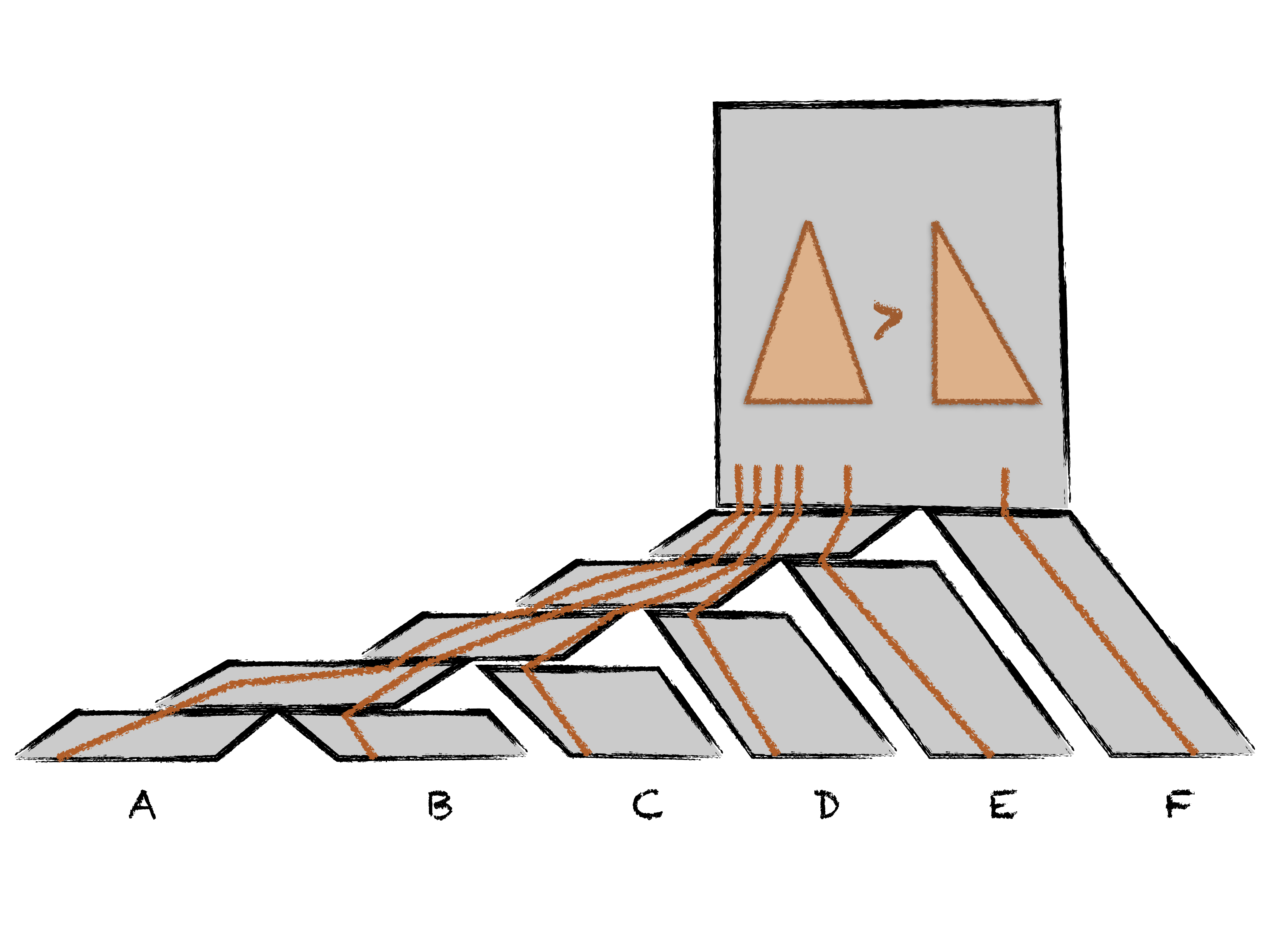}
\caption{Anomalous gene trees on a $6$-taxon species tree
with shape $Z$. The event of deepest coalescence is depicted.}
\label{fig2}
\end{figure}

To establish Theorem~\ref{main} it suffices to do so for any number $n$ of taxa, and we do so for $n=6$.
For the species tree $\species$, take any rooted tree that has the unrooted topology of the $Z$-shaped tree (caterpillar). Make all the edges $L$ of this tree less than $\beta$. 
We use the following notation:
\begin{itemize}
\item Denote by $G_1, \ldots, G_g$
the gene trees generated by the multispecies coalescent
on $\species$.
\item Let $\E_\species$ denote the expectation under $\species$
and let $G$ be a gene tree generated under $\species$.
\item Let $\Xi = [r]^n$ be the set of $r$-state characters
on the set of $n$ taxa.
\item Let $\chi_{k}^f \in \Xi$ 
be the $k$-th character of the
$f$-th alignment, where $1 \leq k \leq \ell$
and $1 \leq f \leq g$, and let $\allchi = \{\chi_k^f\}_{k,f}$. 
\item For a character $\chi \in \Xi$,
let $N_\chi^f$ be the number of times character
$\chi$ appears in the $f$-th alignment and let
$N_\chi$ be the number of times it appears
overall. 
\end{itemize}
Let $U$ be an $n$-leaf tree with mutation probabilites
$\{q_e\}$.
We denote by $p_\chi^U$ the probability that
$\chi$ is produced by $U$ under the symmetric $r$-state
site substitution model.
Then the 
(mis-specified, i.e., not taking into account the
coalescent) empirical minus
log-likelihood under tree $U$ is given by
$$
\loglike_U(\allchi) 
= - \frac{1}{g\ell}\sum_{k,f} \log \left[rp_{\chi_k^f}^U\right]
= - \frac{1}{g\ell}\sum_{\chi \in \Xi} N_\chi \log \left[r p_{\chi}^U\right].  
$$
We want to show that with high probability 
$\loglike_U(\allchi)$ is {\em not} minimized on the species 
tree topology. We follow the proof sketched in Section~\ref{section:preliminaries}.

For a binary tree topology $W$ and a character $\chi \in \Xi$ we let
$\bar{\chi}^W$ denote a minimal extension of $\chi$
on $W$ and $\pars_W(\chi)$, the parsimony score
of $\chi$. Let 
$$
\pars_W(\allchi) = \frac{1}{g \ell}\sum_{k,f} \pars_W(\chi_{k,f})
= \frac{1}{g\ell} \sum_{\chi \in \Xi} N_\chi \pars_W(\chi).
$$
Let $E(W)$ and $V(W)$ be the edges and vertices of $W$.
We assume that $W$ is binary and has $n$ leaves,
hence $|E(W)| = 2n-3$ and $|V(W)| = 2n-2$. 
Let $\loglike_W^*(\allchi)$ be the minus log-likelihood
under an optimal choice of branch lengths (in $[0,+\infty]$) 
for $W$.
Let $N_0$ denote the number of constant characters
and $N_{\neq 0} = g\ell - N_0$.
\begin{claim}[Parsimony-based approximation of the likelihood]
\label{claim:approx-likelihood}
If
\begin{equation}
\label{eq:approx-likelihood-condition}
N_0 > 1, \qquad \frac{g\ell \pars_W(\allchi)}{N_0} \leq 1
\end{equation}
then, for all $q_0 \in (0,1)$,
\begin{equation}
\label{eq:pars-approx-ub}
\loglike_W^*(\allchi)
\leq - \pars_W(\allchi) 
\log\left(\frac{q_0}{r-1}
\right) 
- 2n \log \left(1 - q_0\right)
\end{equation}
and
\begin{equation}
\label{eq:pars-approx-lb}
\loglike_W^*(\allchi)
\geq - \pars_W(\allchi) 
\log\left(
\frac{g\ell \pars_W(\allchi)}{(r-1)N_0}
\right) 
- \frac{N_{\neq 0}}{g\ell} n \log r.
\end{equation}
\end{claim}
{\em Proof:}
We adapt several bounds derived in~\citet[Lemmas 5 and 6]{tuf97}.
Letting $U$ have topology $W$ with all transition probabilities 
equal to $q_0$, by considering a minimal extension
(see~\citet[Equation (52)]{tuf97})
we have 
\begin{eqnarray*}
r p_\chi^U 
&\geq& \left(\frac{q_0}{r-1}\right)^{\pars_W(\chi)}
\left(1- q_0\right)^{2n-3-\pars_W(\chi)}\\
&\geq& \left(\frac{q_0}{r-1}\right)^{\pars_W(\chi)}
\left(1- q_0\right)^{2n},
\end{eqnarray*}
and therefore
\begin{eqnarray*}
\loglike_W^*(\allchi)
&\leq& - \frac{1}{g\ell} 
\left\{
\sum_{\chi \neq 0} N_\chi
\log\left[
\left(\frac{q_0}{r-1}\right)^{\pars_W(\chi)}
\left(1- q_0\right)^{2n}
\right]
+ N_0 \log\left[\left(1-q_0\right)^{2n}\right]
\right\}\\
&=& 
- \pars_W(\allchi)
\log \left(\frac{q_0}{r-1}\right)
- 2n \log (1-q_0),
\end{eqnarray*}
where we used that $N_0 + \sum_{\chi\neq 0} N_\chi = g\ell$.
This proves~\eqref{eq:pars-approx-ub}.

For the other direction,
let $U$ be the tree with topology $W$
and optimal mutation probabilities
$(q_e^*)_e$. Let $\bar{q} = \max_e q_e$.
Then, summing over all minimal extensions
(see~\citet[Equation (63)]{tuf97}),
$$
r p_\chi^U 
\leq r^{n-2} \left(\frac{\bar{q}}{r-1}\right)^{\pars_W(\chi)}
\leq r^{n} \left(\frac{\bar{q}}{r-1}\right)^{\pars_W(\chi)},
$$
and by considering two leaves
whose connecting path goes through
an edge with probability $\bar{q}$ (see~\citet[Equation (9)]{tuf97})
$$
p_0^U \leq 1 - \bar{q}.
$$
Hence
\begin{eqnarray*}
\loglike_U(\chi)
&\geq& -\frac{1}{g\ell}
\left\{
\sum_{\chi \neq 0} N_\chi
\log\left[
r^{n} \left(\frac{\bar{q}}{r-1}\right)^{\pars_W(\chi)}
\right]
+ N_0
\log(1-\bar{q})
\right\}\\
&\geq& 
- \frac{N_{\neq 0}}{g\ell} n\log r
- \pars_W(\allchi) \log \left(\frac{\bar{q}}{r-1}\right) 
+ \frac{N_0}{g\ell}
\bar{q},
\end{eqnarray*}
where we used $-\log(1-\bar{q}) \geq \bar{q}$.
Minimizing $\loglike_U(\chi)$ 
over $\bar{q}$ (see~\citet[Equation (65) and (66)]{tuf97}),
a lower bound is obtained by fixing $\bar{q}$
to $g\ell\pars_W(\allchi)/N_0$.

\hfill$\Box$

In order for the approximation in Claim~\ref{claim:approx-likelihood}
to be useful, we need that $N_0$ is asymptotically larger
than $\max\{n N_{\neq 0},r\}$ and 
that $\pars_W(\allchi)$ is not
too small. We proceed to prove that these two properties hold
when the mutation rate is low enough.

We begin by showing that the empirical frequencies
of characters are close to their expectation when
$g \to +\infty$.
\begin{claim}[Concentration of empirical frequencies]
\label{claim:empirical}
With probability exceeding $1-2r^n \exp(-2g \zeta_1^2)$, for all $\chi \in \Xi$,
\begin{equation}
\label{eq:concentration-freq}
 \left|\frac{1}{g\ell}N_\chi - \E_\species[p_\chi^G]\right| < \zeta_1.
\end{equation}
\end{claim}
{\em Proof:}
For all $\chi \in \Xi$,
\begin{equation}
\label{eq:nchi}
\frac{1}{g\ell}N_\chi 
= \frac{1}{g \ell} \sum_{k,f} \ind_{\{\chi_k^f = \chi\}}
= \frac{1}{g} \sum_{f} \frac{1}{\ell} N_\chi^f
= \frac{1}{g} \sum_{f} \left(\frac{1}{\ell} \sum_{k} 
\ind_{\{\chi_k^f = \chi\}} \right).
\end{equation}
Noting that the $\ell^{-1} N_\chi^fs$ are in $[0,1]$ and independent,
Hoeffding's inequality implies for all $\zeta > 0$
$$
\P_\species\left[\left|\frac{1}{g\ell}N_\chi - \frac{1}{g\ell}\E_\species[N_\chi]\right| \geq \zeta_1 \right] \leq 
2 \exp\left(- 2 g \zeta_1^2\right).
$$ 
Moreover by Eqn.~\eqref{eq:nchi}
\begin{equation*}
\frac{1}{g\ell}\E_\species[N_\chi]
= \frac{1}{g \ell} \sum_{k,f}\E_\species[\ind_{\{\chi_k^f = \chi\}}]
= \P_\species[\chi_k^f = \chi]
= \E_\species[p_\chi^G].
\end{equation*}
The result follows from the fact that $|\Xi| = r^n$.

\hfill$\Box$

An immediate corollary is the concentration of the parsimony
score.
\begin{claim}[Concentration of parsimony score]
\label{claim:concentration-parsimony}
Under Eqn.~\eqref{eq:concentration-freq},
$$
\left|
\pars_W(\allchi)
- \E_\species[\pars_W(\chi)]
\right|
\leq
n r^n \zeta_1.
$$
\end{claim}
{\em Proof:}
By definition,
\begin{eqnarray*}
\left|\pars_W(\allchi)
- \E_\species[\pars_W(\chi)]
\right|
&=& 
\left|\frac{1}{g\ell}\sum_{\chi} \pars_W(\chi) N_\chi
- \sum_\chi \pars_W(\chi) \E_\species[p_\chi^G]
\right|\\
&\leq&
\sum_{\chi} \pars_W(\chi) \left|\frac{1}{g\ell}N_\chi
- \E_\species[p_\chi^G]\right|\\
&\leq&
r^n n \zeta_1.
\end{eqnarray*}
\hfill$\Box$

The next two claims relate the multispecies coalescent
to the standard coalescent. We will refer to 
the population of $\species$ ancestral to all taxa as the {\em master population}.
We let $\deepest$ 
be the gene tree event that no coalescence occurs
before the master population, which we refer to
as {\em deepest coalescence}. We let $\species_\deepest$
be the coalescent model on the master population
(i.e., the standard $n$-coalescent).
We further let $\deepest'$ be the site event
such that $\deepest$ occurs and further
no mutation occurs below the master population.
Let $M$ be the number of mutations on a site. 
\begin{claim}[Lower bound on the number of constant characters]
\label{claim:lower-bound-n0}
There is $\zeta_2$
(depending only on $n$ and $\beta$)
such that,
for any $\theta > 0$, 
$$
\E_\species[p_0^G]
\geq 1-\zeta_2 \theta.
$$
\end{claim}
{\em Proof:}
Note that
$$
\{\chi = 0\} \supseteq \{M=0\}.
$$
The number of mutations on a site is stochastically
dominated by the same quantity {\em conditioned on
$\deepest$}. Indeed deepest coalescence ensures
the highest total length of the gene tree. 
Hence
\begin{equation*}
\E_\species[p_0^G]
\geq \P_\species[M=0]
\geq \P_\species[M=0\,|\,\deepest]
= \E_\species\left[\exp\left(-\theta H_G\right)\,|\,\deepest\right]
\geq \E_\species\left[1-\theta H_G\,|\,\deepest\right],
\end{equation*}
where $H_G$ is the total length of gene tree $G$.
Note that, on $\deepest$,
$$
H_G \leq n\cdot n\beta + H'_G,
$$
where $H'_G$ is the total length of the gene tree 
inside the master population. Letting $h^{(1)}_n$ be the
expected length of the standard coalescent on
$n$ samples, we have
\begin{equation*}
\E_\species[p_0^G]
\geq 1 - \theta (n^2 \beta + h^{(1)}_n).
\end{equation*}
Therefore we can take
$\zeta_2 = n^2 \beta + h^{(1)}_n$.

\hfill$\Box$

\begin{claim}[Reduction to standard coalescent]
\label{claim:reduction-coalescent}
For any $\theta$ and $\zeta_3 > 0$, there is $\beta$ 
small enough
(depending only on $\zeta_3$, $n$, and $\theta$),
such that
$$
\left|
\E_\species\left[\pars_W(\chi)\right]
- \E_{\species_\deepest}\left[\pars_W(\chi)\right]
\right| \leq \zeta_3.
$$
\end{claim}
{\em Proof:}
Note that 
$$
\E_\species\left[\pars_W(\chi)\,|\,\deepest'\right]
= \E_{\species_\deepest}\left[\pars_W(\chi)\right].
$$
Further
\begin{eqnarray*}
\E_\species\left[\pars_W(\chi)\right]
&=&
\E_\species\left[\pars_W(\chi)\,|\,\deepest'\right]
\P_\species[\deepest']
+ \E_\species\left[\pars_W(\chi)\,|\,(\deepest')^c\right]
\P_\species[(\deepest')^c]\\
&\leq&
\E_{\species_\deepest}\left[\pars_W(\chi)\right]
+ n\frac{\zeta_3}{n},
\end{eqnarray*}
by choosing $\beta$ small enough to make the
probability 
$$
\P_\species[\deepest']
\geq (e^{-\binom{n}{2}\beta})^n
\exp(-\theta(n\cdot n\beta))
\geq 1-\frac{\zeta_3}{n}.
$$
Above we used that $\pars_W(\chi) \leq n$.
Similarly,
\begin{eqnarray*}
\E_\species\left[\pars_W(\chi)\right]
&=&
\E_\species\left[\pars_W(\chi)\,|\,\deepest'\right]
\P_\species[\deepest']
+ \E_\species\left[\pars_W(\chi)\,|\,(\deepest')^c\right]
\P_\species[(\deepest')^c]\\
&\geq&
\E_{\species_\deepest}\left[\pars_W(\chi)\right]
\left(1-\frac{\zeta_3}{n}\right)\\
&\geq&
\E_{\species_\deepest}\left[\pars_W(\chi)\right]
- n\frac{\zeta_3}{n}.
\end{eqnarray*}

\hfill$\Box$

Recall that $\E_{\esf}$ is the expectation under the infinite-alleles
model on $\species_\deepest$. 
\begin{claim}[Infinite-alleles approximation]
\label{claim:infinite-approximation}
There is $\zeta_4$ depending
only on $n$ such that,
for any $\theta > 0$, 
$$
\left|
\E_{\species_\deepest}[\pars_W(\chi)]
- \E_{\esf}[\pars_W(\chi)]
\right|
\leq \zeta_4 \theta^2.
$$
\end{claim}
{\em Proof:}
Note that
$$
\E_{\species_\deepest}[\pars_W(\chi)\,|\,M \leq 1]
=
\E_{\esf}[\pars_W(\chi)\,|\,M \leq 1],
$$
as a single mutation as the same effect on the characters
of $r$-state symmetric and infinite-alleles models.
Moreover, because both models are run with the same
parameters, they have the same distribution
of number of mutations. In particular,
$$
\P_{\species_\deepest}[M \leq 1]
= \P_{\esf}[M \leq 1].
$$
Note that
\begin{eqnarray*}
\P_{\esf}[M > 1]
&=&
\E_\esf\left[
\sum_{i\geq 2}
e^{-\theta H_G} \frac{(\theta H_G)^i}{i!}
\right]\\
&\leq&
\E_\esf\left[
(\theta H_G)^2
\sum_{i\geq 0}
e^{-\theta H_G} \frac{(\theta H_G)^i}{i!}
\right]\\
&=& \theta^2 h^{(2)}_n, 
\end{eqnarray*}
where $h^{(2)}_n = \E_\esf[H_G^2]$.
Hence, since
\begin{eqnarray*}
\E_{\species_\deepest}[\pars_W(\chi)]
&=& \E_{\species_\deepest}[\pars_W(\chi)\,|\,M\leq 1]\P_{\species_\deepest}[M \leq 1]\\
&&\qquad+ \E_{\species_\deepest}[\pars_W(\chi)\,|\,M > 1]\P_{\species_\deepest}[M > 1],
\end{eqnarray*}
we have on the one hand
\begin{eqnarray*}
\E_{\species_\deepest}[\pars_W(\chi)]
&\leq& \E_{\esf}[\pars_W(\chi)\,|\,M\leq 1]\P_{\esf}[M \leq 1]\\
&&\qquad + (\E_{\esf}[\pars_W(\chi)\,|\,M > 1] + n)
\P_{\esf}[M > 1]\\ 
&=& \E_{\esf}[\pars_W(\chi)] + n\theta^2 h^{(2)}_n.
\end{eqnarray*}
And, on the other hand, we have
\begin{eqnarray*}
\E_{\species_\deepest}[\pars_W(\chi)]
&\geq& \E_{\esf}[\pars_W(\chi)\,|\,M\leq 1]\P_{\esf}[M \leq 1]\\
&&\qquad + (\E_{\esf}[\pars_W(\chi)\,|\,M > 1] - n)
\P_{\esf}[M > 1]\\ 
&=& \E_{\esf}[\pars_W(\chi)] - n\theta^2 h^{(2)}_n.
\end{eqnarray*}

\hfill$\Box$

\begin{claim}[Final argument]
\label{claim:final}
Let $\theta = \epsilon$. There are
$\epsilon$ and $\beta$ 
small enough (depending on $n$ and $r$)
such that
$
\loglike_Z^*(\allchi)
> \loglike_Y^*(\allchi),
$
with probability exceeding $1-2r^n \exp(-2g \epsilon^4)$.
\end{claim}
{\em Proof:}
Choosing $\beta$ small enough,
$
\zeta_1 = \zeta_3 = \epsilon^2.
$
Claims~\ref{claim:empirical} 
and~\ref{claim:lower-bound-n0}
imply that,
with probability exceeding $1-2r^n \exp(-2g \zeta_1^2)$,
\begin{equation}
\label{eq:lower-bound-n0}
N_0 \geq g\ell[1-\zeta_2\epsilon-\epsilon^2] = g\ell (1-O(\epsilon)),
\qquad N_{\neq 0} = O(g\ell \epsilon). 
\end{equation}
When Eqn.~\eqref{eq:lower-bound-n0} holds,
by Claims~\ref{claim:concentration-parsimony},
~\ref{claim:reduction-coalescent}
and~\ref{claim:infinite-approximation}, 
\begin{equation}
\label{eq:parsimony-approximation}
\left|
\pars_W(\allchi)
-\E_\esf[\pars_W(\chi)]
\right|
=O(\epsilon^2).
\end{equation}
Together with Proposition~\ref{TT'}, this implies that
\begin{equation}
\label{eq:diff-parsimony}
\pars_Z(\allchi)
-\pars_Y(\allchi)
= \frac{1}{60} \epsilon + O(\epsilon^2).
\end{equation}
We finally return to the likelihood. 
Note that~\eqref{eq:approx-likelihood-condition}
in Claim~\ref{claim:approx-likelihood}
is satisfied by~\eqref{eq:lower-bound-n0}
and 
\begin{equation}
\label{eq:final-1}
\pars_Y(\allchi) \leq n N_{\neq 0}/g\ell
= O(n\eps).
\end{equation}
Hence, taking
\begin{equation}
\label{eq:final-2}
q_0 = g\ell\pars_Y(\allchi)/N_0
= O(n\eps),
\end{equation}
in Claim~\ref{claim:approx-likelihood}
yields
\begin{eqnarray*}
\loglike_Z^*(\allchi)
- \loglike_Y^*(\allchi)
&\geq&
- [\pars_Z(\allchi) 
- \pars_Y(\allchi)] 
\log 
\left(
\frac{g\ell \pars_Y(\allchi)}{(r-1)N_0}
\right)\\
&& \qquad + 2n \log \left(1 - 
\frac{g\ell\pars_Y(\allchi)}{N_0}
\right)
- \frac{N_{\neq 0}}{g\ell} n \log r\\
&\geq&
\left[\frac{1}{60} \epsilon + O(\epsilon^2)\right] 
\log [\Omega((n\epsilon)^{-1})]\\
&& \qquad - 2n \cdot O(n\epsilon)
- n \log r \cdot O(\epsilon)\\
&>& 0,
\end{eqnarray*}
by~\eqref{eq:final-1} and~\eqref{eq:final-2},
when $\epsilon$ is small enough
(depending on $n$ and $r$).

\hfill$\Box$

Theorem~\ref{main} now follows immediately from Claim~\ref{claim:final} by noting that the lower bound 
$1-2r^n \exp( -2g\epsilon^4)$ converges to 1 as $g$ grows; consequently, the probability that $Y$ has a higher likelihood than $Z$ (i.e. a lower minus log-likelihood) converges to $1$ 
as the number of alignments $g$ increases.

\section{Concluding comments}
Our statistical inconsistency result applies for the particular case of a tree with six leaves. While this suffices to establish inconsistency in general, we conjecture that an extension of our argument would apply to a tree with any number of leaves. However a detailed proof of this assertion is beyond the scope of this short note.  

\section{Acknowledgements}
The authors thank the Simons Institute at UC Berkeley, where this work was carried out.
S.R.  is supported by NSF grants DMS-1248176 and DMS-1149312 (CAREER), and an Alfred P. Sloan Research Fellowship.
M.S. would like to thank the NZ Marsden Fund and the Allan Wilson Centre for funding support.

\bibliographystyle{plainnat}
\bibliography{concat}
\end{document}